\documentclass[prb,floatfix,twocolumn]{revtex4}

\pdfoutput=1

\usepackage{stmaryrd}

\usepackage{comment}
\usepackage{pdfpages}
\usepackage{graphicx} 
\usepackage{dcolumn} 
\usepackage{amsmath}
\usepackage{amssymb}
\usepackage{bm}
\usepackage{times}
\usepackage{ulem}
\usepackage[justification=raggedright]{caption}

\usepackage{boondox-cal}
\usepackage{subfigure}
\usepackage{tabularx}
\usepackage{appendix}
\usepackage{amstext}
\usepackage{array}

\usepackage{bbm}
\usepackage{blkarray}

\definecolor{wine-stain}{rgb}{0.5,0,0} 
\definecolor{bblue}{rgb}{0,0.0,0.5} 

\usepackage[colorlinks=true
,urlcolor=blue
,anchorcolor=blue
,citecolor=blue 
,filecolor=blue
,linkcolor=wine-stain
,menucolor=blue
,pagecolor=blue
,linktocpage=true
,pdfproducer=medialab
,linktoc=all 
]{hyperref}

\usepackage{booktabs}  

\allowdisplaybreaks

\newcommand{\ncmd}{\newcommand}

\ncmd{\lt}{\left}
\ncmd{\rt}{\right}
\ncmd{\tr}[1]{~\mbox{tr}\lt\{ {#1}\rt\}}
\ncmd{\half}{\frac{1}{2}}
\ncmd{\eps}{\epsilon}
\ncmd{\veps}{\varepsilon}
\ncmd{\dgr}{\dagger}
\ncmd{\sig}{\bar \sigma}
\ncmd{\gam}{\gamma}
\ncmd{\rtarw}{\rightarrow}
\ncmd{\Rt}{\Rightarrow}
\ncmd{\abs}[1]{\lt\cb{#1}\rt\cb}
\ncmd{\avg}[1]{\lt\lb{#1}\rt\rb}
\ncmd{\sgn}[1]{\mbox{sgn}\lt(#1\rt)}
\ncmd{\kap}{\kappa}
\ncmd{\wtil}[1]{\widetilde{#1}}
\ncmd{\thrfr}{\therefore}
\ncmd{\eq}[1]{Eq. \eqref{#1}}
\ncmd{\fig}[1]{Fig. \ref{#1}}
\ncmd{\ordr}[1]{\mathcal{O}\lt(#1\rt)}
\ncmd{\dsty}{\displaystyle}
\ncmd{\alert}[1]{\color{red}{#1}}
\ncmd{\mc}{\mathcal}
\ncmd{\mbf}[1]{\mathbf{#1}}
\ncmd{\Deriv}[2]{\frac{d{#1}}{d{#2}}}
\ncmd{\ParDeriv}[2]{\frac{\partial{#1}}{\partial{#2}}}
\ncmd{\step}[1]{\Theta\lt(#1\rt)}
\ncmd{\td}{\tilde} 
\ncmd{\what}{\widehat}

\ncmd{\ha}{\hat \alpha} 
\ncmd{\hp}{\hat \phi} 
\ncmd{\hpa}{\hat p_\alpha} 
\ncmd{\hpp}{\hat p_\phi} 
\ncmd{\pa}{p_\alpha} 
\ncmd{\pp}{p_\phi} 
\ncmd{\halfd}{\frac{3}{2}} 
\ncmd{\halfdtwo}{\left( \frac{3}{2} \right)^2} 
\ncmd{\halfa}{ \frac{\alpha}{2}} 
\ncmd{\sqEa}{
\sqrt{\kphi ^2+\halfdtwo}
}
\ncmd{\Easqa}{
\frac{ \sqrt{k^2+\halfdtwo}}{3}
} 
\ncmd{\Eatsq}{ 
\frac{\sqrt{E e^{3 \alpha }}}{3} 
}
\ncmd{\Ka}{ K_\alpha} 
\ncmd{\tPsiin}{\Psi_0'(\alpha,\phi) }
\ncmd{\Psiin}{\Psi_0(\alpha,\phi) }
\ncmd{\PsiapD}{\Psi_\CT(\alpha,\phi)}
\ncmd{\PsiapT}{\Psi(\alpha,\phi;\CT)}
\ncmd{\kphi}{q}
\ncmd{\bt}{{\mathcal T}}
\ncmd{\CT}{T}

\ncmd{\BL}{\Bigl\llbracket}
\ncmd{\BR}{\Bigl\rrbracket}

\ncmd{\nn}{\nonumber \\}
\ncmd{\bqa}{\begin{eqnarray}} 
\ncmd{\eqa}{\end{eqnarray}}

\definecolor{new_color}{RGB}{50,155,0}

\ncmd{\vrho}{\varrho}

\makeatletter
\newcommand*{\rom}[1]{\expandafter\@slowromancap\romannumeral #1@}
\makeatother

\begin{document}

\title{
A theory of time
based on wavefunction collapse
}

\author{Sung-Sik Lee\\
\vspace{0.3cm}
{\normalsize{Department of Physics $\&$ Astronomy, McMaster University,}}
{\normalsize{1280 Main St. W., Hamilton ON L8S 4M1, Canada}}
\vspace{0.01cm}\\
{\normalsize{Perimeter Institute for Theoretical Physics,
31 Caroline St. N., Waterloo ON N2L 2Y5, 
Canada}}
}

\date{\today}

\begin{abstract}
We propose that moments of time arise through the failed emergence of the temporal diffeomorphism as gauge symmetry,
and that the passage of time is a continual process of an instantaneous state collapsing toward a gauge-invariant state.
Unitarity and directedness of the resulting time evolution are demonstrated for a minisuperspace model of cosmology.
\end{abstract}

\maketitle

\newpage

\section{Introduction}

Time is the most fundamental concept in physics, yet the least understood one.
In the Newtonian paradigm,
time is a parameter that labels moments of history and endows them with a chronological order.
Like the conductor of an orchestra who silently leads other musicians,
time itself is not observable but provides incessant cues for physical degrees of freedom to march on.
Physical laws dictate how dynamical variables evolve as functions of time, 
but explaining the flow of time is not necessarily a mandate of theories in this framework.

In Einstein's theory of gravity,
the time translation 
is merely a gauge transformation that generates redundant descriptions of one spacetime.
In the absence of an absolute time, specifying a moment without a reference to dynamical variables is impossible 
\cite{PhysRevD.65.124013,Dittrich2007}.
While the theory predicts correlation among physical observables,
it does not explain why events unfold in a particular order.
Therefore, relational theories such as general relativity 
are challenged with 
the tasks of 
finding a dynamical variable that can serve as a clock
and reconciling our experience of instants that persistently pass by with the four-dimensional block universe present once and for all.

Quantizing gravity\cite{PhysRev.160.1113} comes with new challenges related to time
\cite{1992gr.qc....10011I,1992grra.conf..211K,https://doi.org/10.1002/andp.201200147}. 
Here, we focus on one.
Suppose  $|\Psi\rangle$ is a state that is invariant under the temporal diffeomorphism.
Because $\hat H |\Psi\rangle=0$, 
where $\hat H$ is the generator of the temporal diffeomorphism, 
the dynamical information is solely encoded in the entanglement of physical degrees of freedom\cite{PhysRevD.27.2885}.
A moment is defined through a measurement of a variable chosen as a clock.
The entanglement between the clock and other variables determines the dynamics, that is, the latter's dependence on the former.
However, there are many ways of defining moments, even for one clock variable,
because there is in prior no preferred basis in which the clock variable should be measured.
In general, a rotation of the basis of clocks defines different moments of time
and can even alter the notion of locality in space if the rotation creates non-trivial entanglement between local clocks\cite{Lee:2021ta}.
In appendix 
\ref{app:txexample},
we discuss a simple example that illustrates this.

The fundamental difficulty of defining time in relational quantum theories is that the notion of instant is not gauge invariant.
No matter what clock we choose, the state of an instant that arises from a projective measurement of the clock is not gauge invariant.
Therefore, restoring time in quantum gravity may involve reconsidering the role of the temporal diffeomorphism as gauge symmetry\cite{
PhysRevD.79.084008,
2008arXiv0808.1223B}.
For other ideas on the origin of time,
see Refs. \cite{
time_timeless,
Horwitz:1988aa,
ISHIBASHI1997467,
Connes:1994aa,
SMOLIN201586,
Carroll:2004pn,
MAGUEIJO2021136487,
Hull:2014aa,
Lee:2020aa,
PhysRevD.108.086020,
Brahma:2022aa}

\section{Emergent time from a collapse of wavefunction}

\begin{figure}
    \centering
\includegraphics[width=0.9\linewidth]{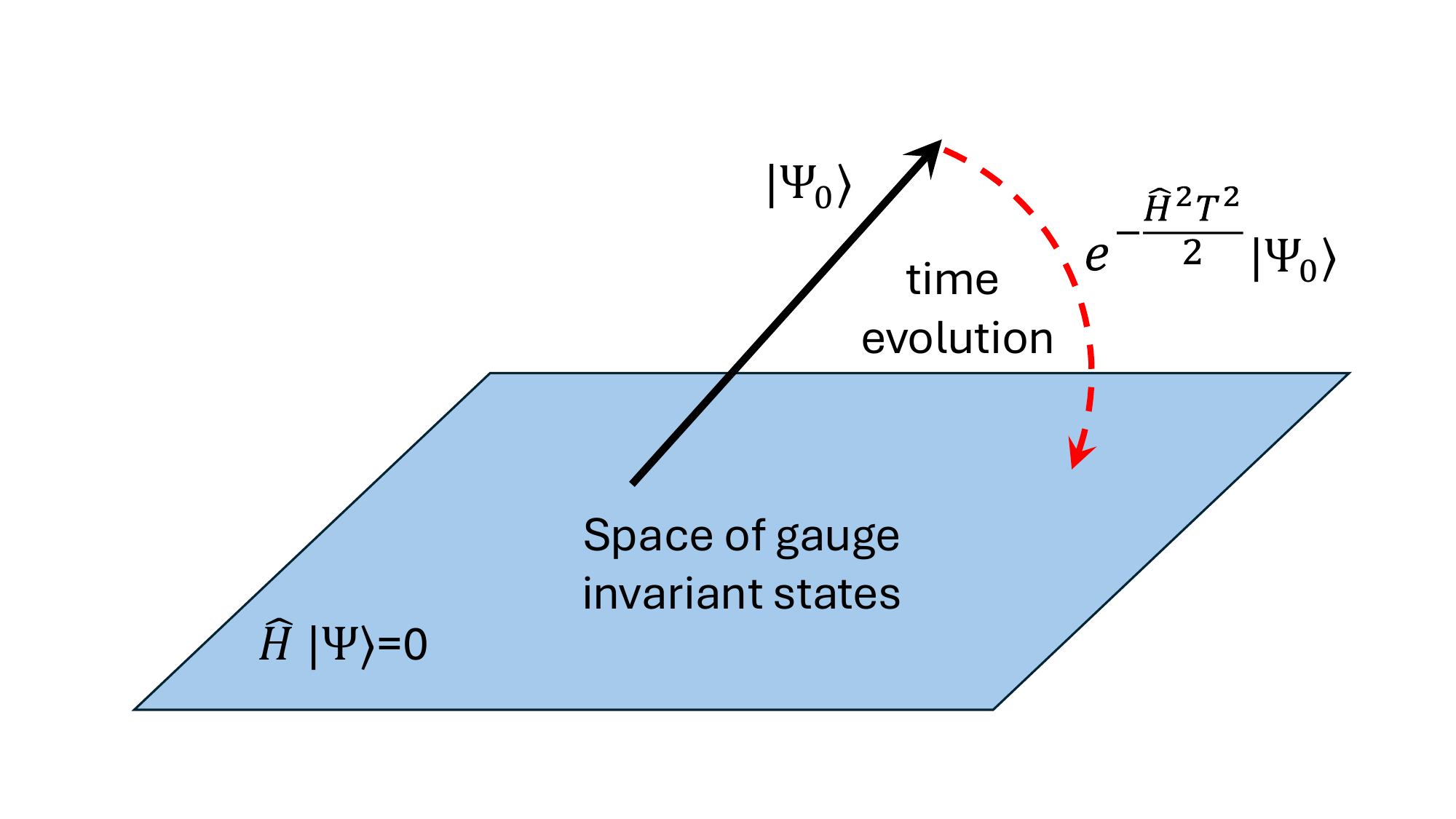}
    \caption{
The continual collapse of a gauge non-invariant initial state toward a gauge invariant state as a time evolution.
}
    \label{fig:collapse}
\end{figure}

In this paper, we posit that the temporal diffeomorphism is not fundamental but only approximate.
This amounts to including states that are not gauge invariant within the physical Hilbert space.
In condensed matter systems where gauge theory emerges at low energies through gauge constraints imposed approximately at the microscopic scale \cite{KITAEV20062,PhysRevLett.90.016803,PhysRevB.69.064404},
such extended Hilbert spaces arise inevitably
because strictly gauge-invariant Hilbert spaces can not be written as a product of local Hilbert spaces\footnote{
Such an extension can be viewed as an inclusion of matter fields.
}. 
In quantum gravity, an extension of Hilbert space is needed to represent moments of time.
With this extension, we make our main proposal:
\begin{enumerate}
    \item 
{\it 
Each moment of time is represented by a quantum state, which is not invariant under the temporal diffeomorphism.
}
\item 
{\it
Time evolution is a continual process in which an initial state collapses toward a gauge invariant state.
}
\end{enumerate}
We now fill in the details of the proposal.
Without loss of generality, an initial state is written as 
$|\Psi_0 \rangle  = \int dE d\kphi  ~
\Phi_0(E,\kphi )
|E,q \rangle$.
Here,
$|E,q \rangle$ is
the eigenstate of 
$\hat H$ with eigenvalue $E$.
$\kphi $ labels the state of the 
 physical degrees of freedom that can be varied within the gauge invariant Hilbert space with $E=0$.
We assume that 
$|\Phi_0(E,\kphi )|^2$ is 
integrable and analytic as a function of $E$.
The latter condition is equivalent to requiring that the wavefunction is exponentially localized in the variable conjugate to $\hat H$ that can be viewed as time.
Because $\Phi(E,q)$ is generally non-zero at $E\neq0$, $|\Psi_0\rangle$ is not gauge invariant.
In our proposal, a continual projection of $|\Psi_0\rangle$ toward a gauge invariant state corresponds to the time evolution.
Such a projection 
can be implemented through a random walk 
within the gauge orbit where
states with $E\neq 0$ are suppressed through a destructive 
interference. 
Here, one step of the random walk is taken by 
 $e^{i \hat H \epsilon}$ 
 or 
 $e^{-i \hat H \epsilon}$,
 where $\epsilon$ is an infinitesimal step size with the randomly chosen sign.
The state obtained from averaging over all paths of $N$ steps becomes 
\bqa
|\Psi_N\rangle
\sim \sum_{\{\epsilon_j\}}
e^{i \sum_{j=1}^N \epsilon_j \hat H}
|\Psi_0\rangle,
\label{eq:averagedstate}
\eqa
where $\epsilon_j= \epsilon$ or $-\epsilon$.
In the large $N$ limit
with fixed $\CT = \sqrt{N} \epsilon$,
the net gauge parameter 
$\tau \equiv \sum_{j=1}^N \epsilon_j$ acquires the Gaussian distribution with width $\CT$,
and
\eq{eq:averagedstate} becomes
$|\Psi(\CT) \rangle=
{\cal N}(\CT) 
\int_{-\infty}^\infty 
d \tau ~ 
e^{-\frac{1}{2\CT^2} \tau^2}
e^{i \hat H \tau}
|\Psi_0\rangle$,
where ${\cal N}(\CT)$ is a normalization.
Integrating over $\tau$, 
one obtains 
\bqa
|\Psi(T)\rangle =
\sqrt{2\pi} \CT
{\cal N}(\CT) 
\int dE d\kphi  
e^{-\frac{\CT^2}{2} E^2}
\Phi(E,\kphi )
|E,q \rangle,
\label{eq:mastereq2}
\eqa
which describes a continuous projection of the initial state toward a gauge-invariant state.
In our proposal, $T$ is {\it time} and \eq{eq:mastereq2} represents the time-dependent quantum state. 
This is illustrated in \fig{fig:collapse}.

In our theory, an initial state is gradually projected toward a gauge invariant state under the time evolution.
Even though $|\Psi(T) \rangle$ at finite $T$ 
is not annihilated by $\hat H$,
the state at every $T$ is within the physical Hilbert space.
With the enlarged Hilbert space,
the present theory has additional observables beyond the gauge-invariant ones.
Since the temporal diffeomorphism is not a strict gauge symmetry,
physical observables do not need to commute with $\hat H$.
The most important additional observable is the conjugate momentum $\hat p_H$ of $\hat H$,
which  does not commute with $\hat H$ by definition.
Our time variable $T$ is closely related to $\hat p_H$ through
$T^2 \sim \frac{1}{\langle \hat H^2 \rangle}
\sim 
\langle \hat p_H^2 \rangle$,
where the first relation is through the definition of our $T$ variable
and the second relation is due to the uncertainty relation between $\hat H$ and $\hat p_H$.

The exact gauge constraint is restored at 
$\CT = \infty$.
However, $|\Psi(T)\rangle$ at any finite $T$ is qualitatively different from a strictly gauge invariant state 
for $\hat H$ that generates a non-compact group.
To quantify the residual violation of the gauge constraint left at time $T$,
we use the normalized trace distance 
$ d_\Psi(y) \equiv
\frac{1}{2
\langle \Psi | \Psi \rangle
} 
tr
\left\{
\left| 
| \Psi \rangle \langle \Psi | 
-
e^{i y \hat H} | \Psi \rangle   \langle  \Psi | e^{-i y \hat H} 
\right|
\right\} $
that measures the distance between $|\Psi\rangle$ and $e^{i y \hat H}|\Psi\rangle$:
for gauge invariant $|\Psi\rangle$, $d_\Psi(y) = 0$ for all $y$;
if $|\Psi\rangle$ and $e^{i y \hat H}|\Psi\rangle$ are orthogonal, 
$d_\Psi(y) = 1$;
otherwise it takes values between $0$ and $1$.
If $\hat H$ is a generator of a compact group, the spectrum of $\hat H$ is discrete.
Because the gauge non-invariant components 
of $|\Psi(\CT)\rangle$ are uniformly suppressed at large $\CT$, 
$|\Psi(\CT)\rangle$ can be made arbitrarily close to a gauge-invariant state:
for any non-zero $\delta$, there exists a sufficiently large $\CT$ such that $d_{\Psi(\CT)}(y) < \delta$ for all $y$.
The situation is different for the non-compact temporal diffeomorphism,
where gauge-invariant states are generally within a band of states with continuously varying eigenvalues.
In such cases, no matter how large $\CT$ is,
there always exists a sufficiently large $y$ such that $d_{\Psi(\CT)}(y)$ is $O(1)$.
This can be seen from the trace distance between 
$|\Psi(\CT) \rangle = 
e^{- \frac{\CT^2}{2} \hat H^2} |\Psi_0\rangle$
and $e^{i\hat H y} |\Psi(\CT)\rangle$,
$d_{\Psi(\CT)}(y) =
\sqrt{
1- 
\left| 
\frac{
\int dE dq ~  |\Phi_0(E,q)|^2 e^{-\CT^2 E^2 + i y E}}
{
\int dE  dq~  |\Phi_0(E,q)|^2 e^{-\CT^2 E^2}}
\right|^2 }
$.
For smooth $|\Phi_0(E,q)|^2$,
$\lim_{y \rightarrow \infty} d_{\Psi(\CT)}(y) = 1$ for any finite $\CT$.
In this sense, the non-compact gauge symmetry does not emerge 
at a finite $T$.
This difference also affects whether the Coulomb phase of a gauge theory can emerge or not in lattice models with soft gauge constraints as is discussed in Appendix  
\ref{app:gauge_symmetry}.

For the temporal diffeomorphism, 
truly gauge invariant states, 
which are stationary against the evolution generated by $\hat H$, are never reached at any finite $T$.
We view this failed emergence of gauge symmetry as the underlying reason why moments of time exist and time continues to flow.
There is a similarity between this and how the bulk space emerges in holographic duals of field theories\cite{
Maldacena:1997re,Witten:1998qj,Gubser:1998bc}.
The renormalization group flow,
which generates the radial direction of the emergent bulk space\cite{
deBoer:1999xf,
Skenderis:2002wp},
can be understood as 
the gradual collapse of a state associated with a UV action toward the state associated with an IR fixed point
through an action-state mapping\cite{Lee:2013dln}.
Here, the UV state, which is not annihilated by the radial constraint, exhibits a non-trivial RG flow,
and
the inability to project a highly entangled UV state to the trivial IR state creates a space with infinite radial depth in the bulk\cite{Lee2016}.

Before we can interpret such wavefunction collapses as time evolution,
however, we have to address two immediate issues.
The first is unitarity.
In general, the projection of a wavefunction causes its norm to change.
One can enforce unitarity by choosing 
${\cal N}(\CT)^{-2} = 2 \pi \CT^2  \int dE d\kphi   |\Phi(E,\kphi )|^2 e^{-\CT^2 E^2}$
so that the norm of 
\eq{eq:mastereq2}
is independent of $\CT$.
The resulting unitary evolution is generally non-linear due to the dependence of ${\cal N}(\CT)$ on the state.
In the large $T$ limit, however, ${\cal N}(\CT)$ only depends on the the  $E \rightarrow 0$ limit of $\int dq |\Phi(E,\kphi)|^2$.
For $\int dq |\Phi(E,\kphi)|^2$ that is analytic at $E=0$, 
distinct classes of initial states are characterized by one even integer $n \geq 0$ that sets the small $E$ limit of the wavefunction as
$\int dq |\Phi(E,\kphi)|^2 \sim E^{n}$. 
For an initial state with exponent $n$,
${\cal N}(\CT) \sim T^{(n-1)/2}$ 
in the large $T$ limit.
Crucially, ${\cal N}(\CT)$ does not depend on the state of the physical degrees of freedom denoted by $q$. 
Consequently, a linear and unitary evolution emerges in the large $\CT$ limit. 
States with different exponents can be thought to be in different superselection sectors in that each state and its late-time dynamics are characterized by single exponent.
\footnote{
This is analogous to the way universality emerges 
as the renormalization group flow is
controlled by a small set of data associated with relevant and marginal couplings at long-distance scales.}
The emergence of linear unitary evolution will be demonstrated through an explicit calculation for the most generic case of $n=0$. 
States with $n>0$,
which can be studied in the similar way,
form a measure-zero set as they require fine-tuning. 
%

%
The second issue is the directedness of time.
The gradual projection of the wave function is the result of a stochastic evolution along the gauge orbit.
Under such an evolution, a state usually diffuses in all directions in the gauge orbit.
If one of the variables is used as a clock, 
the diffusion would create a state that is merely more spread over a more extensive range of past and future 
without pushing time in one direction.
However, a directed time evolution can emerge from such a stochastic evolution if $\hat H$ is asymmetric in the space of configuration.
One such example is general relativity.
In the canonical formulation of general relativity, the Hamiltonian density 
in three space dimensions reads
$
\hat h = 
\frac{1}{ 
\sqrt{g}  
} 
\left(
\Pi^{\mu \nu} \Pi_{\mu \nu}
- \frac{1}{2} \Pi^2 \right) 
- 
\sqrt{g}  
R
$,
where $g=\det{g}$ measures the proper volume of a spatial region with a unit coordinate volume,
$\Pi^{\mu \nu}$ is the conjugate momentum of $g_{\mu \nu}$,
$\Pi \equiv \Pi^{\mu}_{\mu}$
and $R$ is the three-dimensional scalar curvature.
In the kinetic term quadratic in $\Pi$,
the factor of $\sqrt{g}$
can be viewed as the `effective mass' of metric. 
This captures the intuitive fact that the universe becomes `heavier' as its size increases.
Since the dynamics becomes slower at larger $g$,
configurations generated through the random walk at a larger $g$ add up with a stronger constructive interference in the ensemble of
\eq{eq:averagedstate}.
This configuration-dependent effective mass makes the state of the universe evolve preferably toward the one with larger size with increasing $\CT$.
In the following, 
we explicitly demonstrate 
the unitarity and directedness of the time evolution  
for the minisuperspace truncation of general relativity.
However, these features are 
 expected to hold for a broader set of models with configuration-dependent effective mass.

\section{Application to cosmology}
We consider the Friedmann–Robertson–Walker (FRW) model for
the scale factor ($e^{\hat \alpha}$) of a three-dimensional space
and a massless free scalar ($\phi$).
The Hamiltonian reads
\bqa
\hat H = 
\BL
e^{-3 \ha} 
(-\hpa^2 +  \hpp^2 )
+ e^{3 \ha}  \rho(\ha)
\BR,
\label{H_FRW}
\eqa
where $\hpa$ and $\hpp$ are the conjugate momenta of $\ha$ and $\hp$, respectively,
and
$\llbracket \hat O \rrbracket \equiv \frac{1}{2} ( \hat O + \hat O^\dagger)$.
$\hat H$ is symmetric but not essentially self-adjoint for general square integrable wavefunctions due to the divergence at $\alpha=-\infty$.
We consider the $\alpha$-dependent energy density of the form,
\bqa
\rho(\alpha) & = & 
\Lambda_{c}(\alpha)
+ \Lambda_m e^{-3 \alpha}
+ \Lambda_r e^{-4 \alpha}.
\label{eq:rho}
\eqa
Here, 
$\Lambda_{c}(\alpha) =
\Lambda_{0}+
\Lambda_{1} e^{-2 \alpha}$;
$\Lambda_0$ is the $\alpha$-independent cosmological constant,
and 
$\Lambda_1$ includes the component of the dark energy that decays as $e^{-2\alpha}$\cite{OZER1987776, FREESE1987797, PhysRevD.46.2404, PhysRevD.41.695}
and the contribution of the spatial curvature.
Henceforth, $\Lambda_c(\alpha)$ will be simply called the dark energy.
$\Lambda_m$ and $\Lambda_r$ represent the contributions of matter and radiation, respectively.
\eq{H_FRW} can be obtained by projecting 
a Hamiltonian of all degrees of freedom
to a sub-Hilbert space 
in which the degrees 
 of freedom other than $\alpha$ and $\phi$ are in an $\alpha$-dependent state 
 (see Appendix \ref{app:FRW}).
The Planck scale is set to be $1$.



We write eigenstates of 
 \eq{H_FRW} with eigenvalue $E$ as
$\Psi_{E,q}(\alpha,\phi) = e^{i q \phi + \frac{3}{2} \alpha} f_{E,q}(\alpha)$,
where
$f_{E,q}(\alpha)$ satisfies
\bqa
f^{''}_{E,q}(\alpha)
+ P_{E,q}(\alpha)^2
f_{E,q}(\alpha) = 0,
\eqa
where
\bqa
P_{E,q}(\alpha)^2
= 
q^2 + 
(3/2)^2
+  \rho(\alpha) e^{6 \alpha} 
-  E e^{3 \alpha}.
\label{eq:PE}
\eqa
For simplicity, we focus on states with $q \lesssim 1$,
and assume that there exists a hierarchy among different types of energy densities such that  
$(\Lambda_{1}/\Lambda_{0})^{1/2}  \gg
\Lambda_m/\Lambda_{1}  \gg
\Lambda_r/\Lambda_m  \gg
1/\Lambda_r^{1/2}
\gg 1$.
In this case, the evolution undergoes a series of crossovers at
$\alpha_A \sim \log (1/\Lambda_r^{1/2})$,
$\alpha_B \sim \log (\Lambda_r/\Lambda_m)$
and
$\alpha_C \sim \log (\Lambda_m/\Lambda_1)$.
Between these crossover scales,
one of the terms dominates the energy density in \eq{eq:PE},
which results in the following epochs:
1) pre-radiation era 
($\alpha \ll \alpha_{A})$,
2) radiation-dominated era 
($\alpha_{A} \ll \alpha \ll \alpha_{B})$,
3) matter-dominated era,
($\alpha_{B} \ll \alpha \ll \alpha_{C})$,
4) dark-energy-dominated era
($\alpha_{C} \ll \alpha)$.
The dark-energy-dominated era is further divided into two sub-eras around
$\alpha^* \sim \frac{1}{2} \log ( \Lambda_1/\Lambda_0)$,
depending on whether the $\Lambda_1$ or $\Lambda_0$ term dominates the dark energy.
Below, we describe the evolution of the universe in each era.

Within the mini-superspace framework, 
a fixed non-zero $\hat H$,
where $\hat H$ represents the spatial integration of the Hamiltonian density of general relativity, 
would look like a cold matter.
This can be seen from Eqs.
\eqref{H_FRW}
and
\eqref{eq:rho},
where a non-zero $\hat H$ shifts the total energy from matter, $\Lambda_m$.  
However, its effect on dynamics is different from that of cold matter because
the average value of  $\langle \hat H^2 \rangle$ is not a constant but decreases with increasing $T$.

\subsection{Pre-radiation era}

In this era, the Hamiltonian constraint becomes
${\cal H} = 
\BL
e^{-3 \alpha} 
\left( \partial_\alpha^2 - \partial_\phi^2 \right)
\BR$.
This may not describe the realistic pre-radiation era as it ignores other effects, such as inflation.
Nonetheless, we study this as a toy model because the exact solution available in this limit is useful for demonstrating the general idea without an approximation.
Normalizable eigenstates of ${\cal  H}$ have non-positive eigenvalues.
Eigenstates with eigenvalue $E$ 
($E \leq 0$),
which are regular in the small $|E|$ limit,
are given by
\bqa
\Psi_{E,\kphi }^{(\pm)}(\alpha,\phi) &=& 
e^{\halfd \alpha+  
i \kphi  \phi}  
(-E)^{ \mp  i \frac{\epsilon_{\kphi }}{3} } ~ J\left[ \pm i  \frac{2}{3} \epsilon_{\kphi } ; \frac{2}{3}  \sqrt{-E} e^{\halfd \alpha} \right], ~~~~~~
\label{eq:generalsol}
\eqa
 where 
 $J[\nu;z]$ is the Bessel function of the first kind of order $\nu$
 and
$\epsilon_{\kphi } = \sqrt{\kphi ^2 + \halfdtwo} $.
In the small $|E|$ limit, \eq{eq:generalsol} reduces to gauge-invariant states: 
$\lim_{E\rightarrow0}\Psi_{E,\kphi }^{(\pm)}(\alpha,\phi)  \sim 
e^{\halfd \alpha}  
e^{i \kphi  \phi \pm i \epsilon_{\kphi } \alpha}$.
Because the amplitude of the gauge invariant state grows exponentially in $\alpha$,
a projection of $|\Psi_0\rangle$ with finite support in $\alpha$ toward such gauge invariant states is expected to make the state evolve toward the region of large $\alpha$ to maximize the overlap. 
A general normalizable gauge non-invariant state can be written as
\bqa
 \Psi_{0}(\alpha,\phi)
&= &
\sum_{s=\pm} \int_{-\infty}^0 dE \int d\kphi 
~
\Phi_s(E,\kphi ) \Psi_{E,\kphi }^{(s)} (\alpha,\phi).~~
\label{PsiEk_general}
\eqa
For $\Phi_{\pm}(E,\kphi )$ 
that is smooth in $E$,
\eq{eq:mastereq2}
at large $\CT$ becomes
\bqa
\Psi(T)
=
\sum_{s=\pm}
\int d\kphi ~
\Phi_s(0,\kphi )
e^{i \kphi  \phi  + i s  \epsilon_{\kphi } \alpha }
\chi_{\kphi ,s} \left(
\frac{e^{\alpha}}{\CT^{1/3}} 
\right) 
\label{psiaphiDnonchiral}
\eqa
up to terms that vanish as $1/T$,

where
$
\chi_{\kphi ,\pm}(z)  =
6^{ \mp i\frac{2\epsilon_{\kphi }}{ 3}}  
\frac{\pi^{1/2}}{2} 
z^{\frac{3}{2}}  
\Bigg[ ~
 _0\tilde{F}_2\left(; 
 \frac{3 \pm 2 i\epsilon_{\kphi }}{6}, 
\frac{3 \pm i \epsilon_{\kphi }}{3}; 
\frac{z^6}{648}
  \right) $
$ -
\frac{\sqrt{2}}{36} ~
_1\tilde{F}_3\left( 1; \frac{3}{2}, 
\frac{3 \pm i \epsilon_{\kphi }}{3}, 
 \frac{9 \pm 2 i\epsilon_{\kphi }}{6}; 
  \frac{z^6}{648}
\right)z^3
\Bigg]$
with
$_p\tilde{F}_{p'}\left({a_1,..,a_p};{b_1,..,b_{p'}};x \right)
=
\frac{  
_p F_{p'} \left({a_1,..,a_p};{b_1,..,b_{p'}};x \right)}{
\Gamma(b_1)..\Gamma(b_{p'}) }$.

\begin{figure}
    \centering
\includegraphics[width=0.9\linewidth]{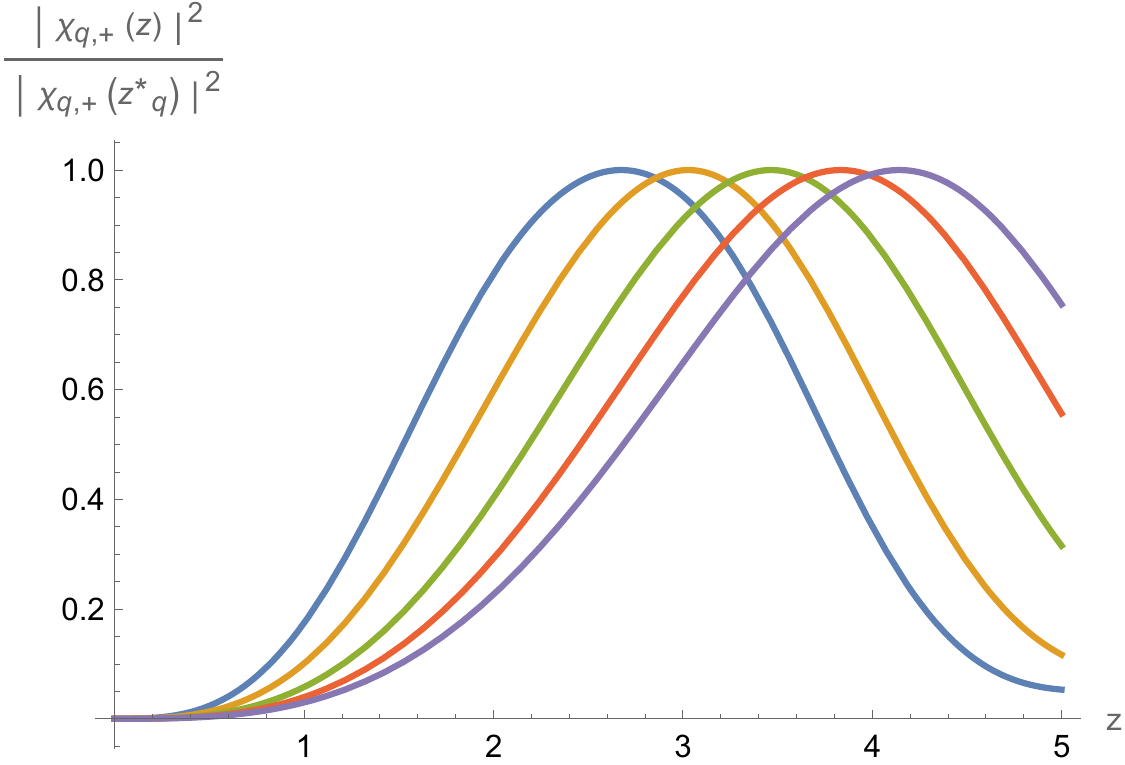}
    \caption{
$|\chi_{\kphi ,\pm}(z)|^2$ 
normalized by its peak value at $z_{\kphi}^*$ 
with increasing
$q=0,2,4,6,8$ from left to right curves.
The scale factor at which the wavefunction is peaked increases with $\kphi$ because a larger momentum of $\phi$ gives rise to a larger momentum for $\alpha$. 
}
    \label{fig:chikz}
\end{figure}

\eq{psiaphiDnonchiral}
describes the evolution of the state 
as it gradually collapses toward a gauge invariant state with increasing $\CT$
which is regarded as 
{\it time}.
$\chi_{\kphi ,\pm}(z)$,
which controls the magnitude of the wavefunction for each component of $q$,
is peaked at $z_{\kphi }^*$, 
as is shown in \fig{fig:chikz}.
The wavefunction for $\alpha$ is peaked at $\CT$-dependent $\alpha(\CT)$ with a finite uncertainty.
At time $\CT$,
$e^{\alpha(\CT)} \sim \CT^{1/3}$,
and
its conjugate momentum is 
$p_\alpha(\CT) \approx \pm \epsilon_q$.
While $\PsiapT$ is not gauge invariant,
$\langle \hat \alpha \rangle$, 
$\langle \hat p_\alpha \rangle$, 
$\langle \hat \phi \rangle$, 
$\langle \hat p_\phi \rangle$
satisfy the classical Hamiltonian constraint up to a correction 
that goes to zero in the large $T$ limit.
Since the state of $\alpha$
is fixed at each $\CT$,
$\alpha$ is not an independent dynamical variable. 
The scalar, which retains information about the initial state, is the physical degree of freedom. 
Therefore, the present theory keeps the same number of physical degrees of freedom as the system in which the gauge symmetry is strictly enforced.
For $T \gg 1$,
the norm of the wavefunction is independent of $\CT$ 
and
the resulting 
time evolution can be written as 
$\Psi(\alpha,\phi;\CT+\Delta \CT) = e^{-i \Delta \CT \hat H_{eff}(\CT)} \Psi(\alpha,\phi;\CT)$,
where 
$\hat H_{eff}(\CT)
=
\frac{1}{3 \CT} 
\BL \hpa -
\hat \Pi
\sqrt{\hpp^2+\halfdtwo} \BR$
is the effective Hamiltonian.
Here, $\hat \Pi$ is the operator that takes eigenvalues $\pm 1 $ for   
$\psi^{(\pm)}_{E,\kphi }$.\footnote{
$\hat \Pi$
commutes with 
$\hpp$ because a translation in 
$\phi$ does not change the parity of
$\psi^{(\pm)}_{E,\kphi }$.}
We note that this unitary and linear time evolution is a phenomenon that emerges in the large $T$ limit for generic initial states.
This can be seen from the fact that the effective Hamiltonian is independent of the details of the initial state.
The effective Hamiltonian 
makes $\alpha$ to increase with increasing time 
irrespective of $p_\alpha$.
This arrow of time arises because the preferred direction of the gauge parameter is determined by the state:
for states with 
$p_\alpha > 0$ 
($p_\alpha < 0$),
$e^{i \epsilon \hat H}$
($e^{-i \epsilon \hat H}$)
generates a stronger constructive interference to always push the state to larger $\alpha$.

Within the gauge orbit generated by the Hamiltonian,
the scale factor generally increases in one direction and decreases in the opposite direction.
The standard approaches to quantum gravity do not provide a mechanism that determines the direction of physical time within the gauge orbit. 
In the present theory, the direction of physical time $T$ is dynamically determined such that $\frac{d e^{\alpha}}{dT}$ is positive. 
Accordingly, the theory predicts that space can not shrink forward in time.
Here, we emphasize that $\alpha$, which is bound to increase with $T$, represents the overall size of the universe.
For a spatial manifold that is compact, it determines the proper volume of the entire universe, which is finite and increasing with $T$.
However, it still allows the gravitational collapse of matter that forms structures in the universe.
Namely, $\frac{d \alpha}{dT} > 0$ is compatible with the focusing nature of general relativity that guarantees 
$\frac{d\theta}{dT} \leq 0$ under a positive energy condition, 
where $\theta$ is the speed at which nearby geodesics move apart from each other.
Since $\frac{d\theta}{dT}$ is even under the time reversal,
the acceleration of geodesic congruences is negative along either direction of the gauge orbit. 
Therefore, the attractive nature of gravity is not affected by the lapse function that is dynamically selected in this theory.

\subsection{
Radiation and matter-dominated eras
}

At $\CT_A = 1/\Lambda_r^{3/2}$,
the peak of the wavefunction reaches the first crossover scale:
$\alpha(\CT_A) \sim \alpha_A$.
For $\CT > \CT_A$, the evolution becomes dominated by radiation and then matter consecutively.
We consider the two eras together
because the analysis is parallel for those two cases. 
In each era, we can keep only one dominant term in the energy density to write \eq{eq:PE} as
\bqa
P_{E,q}(\alpha)^2= C_n e^{n \alpha} - E e^{3 \alpha}
\label{PEq23}
\eqa
with $C_2=\Lambda_r$ and $C_3=\Lambda_m$, respectively.
In solving \eq{eq:PE}, it is useful to understand the relative magnitude between the two terms in 
\eq{PEq23} for typical values that $E$ and $\alpha$ take.
At time $\CT$, the range of $E$ in  \eq{eq:mastereq2} is $E(\CT) \sim \CT^{-1}$ while the wavefunction is peaked at $\alpha(\CT)$.
At $\CT_A$, the two terms are comparable:
$\frac{C_2 e^{2 \alpha(\CT_A)}}{ E(\CT_A) e^{3 \alpha(\CT_A)}}
\sim
\Lambda_r e^{2 \alpha_A} \sim 1$\footnote{It follows from the fact that
$\CT \sim e^{3\alpha(\CT)}$ in the pre-radiation era.}.
For $\CT \gg \CT_A$, 
a hierarchy emerges such that
\bqa
C_ne^{n \alpha(\CT)}
\gg E(\CT) e^{3 \alpha(\CT)} \gg 1.
\label{eq:smallE}
\eqa
This will be shown to be true through a self-consistent computation in the following.
For now, we proceed, assuming that this is the case.
With $P_{E,q} \gg 1$, we can use the WKB-approximation to write the eigenstates 
 of ${\cal H}$ with eigenvalue $E$ as
\bqa
\Psi_{E,q}^{(s)}(\alpha,\phi) =
e^{
\frac{3}{2} \alpha
+
i q \phi 
}
\frac{
\exp{\left[  i s \int P_{E,q}(\alpha) d\alpha \right]}
}{ \sqrt{P_{E,q}(\alpha)} }
\label{eq:PsiEqn}
\eqa
with $s=\pm 1$.
Furthermore,  \eq{eq:smallE} allows us to expand 
\eq{eq:PsiEqn} around $E=0$ to write
$\Psi_{E,q}^{(\pm)}(\alpha,\phi) \approx
\frac{ e^{\frac{3}{2} \alpha
+ i  \left[  q \phi  \pm  \int  P_{0,q}(\alpha) d\alpha  \mp   E \int  \frac{e^{3\alpha}}{ 2 P_{0,q}(\alpha)} d\alpha  \right]
}} { \sqrt{P_{0,q}(\alpha)} }$.
To the leading order in $1/T$,
the integration over $E$ 
in \eq{eq:mastereq2} 
leads to
\bqa
\Psi(T) &=&
\sum_{s=\pm}
\int d\kphi ~
\Phi_s(0,\kphi )
e^{
i \left( \kphi  \phi  
+  2 s \frac{\sqrt{C_n}}{n} e^{\frac{n}{2} \alpha } 
 \right) } \times \nn &&
\chi_n \left(
\frac{e^{\alpha}}
{ (C_n \CT^2)^{\frac{1}{6-n}} }
\right), 
\label{psiaphigeneral}
\eqa
where
$
\chi_n(z)  =
z^{\frac{6-n}{4}} 
 e^{ -\frac{z^{6-n}}{2 (6-n)^2}}
$.
At time $\CT$, the wavefunction is peaked at 
$e^{\alpha(\CT)} \sim 
 (C_n \CT^2)^{\frac{1}{6-n}} 
$. 
In the radiation-dominated era,
the size of the universe increases as  
$e^{\alpha(T)} \sim 
e^{\alpha_A}
\left( T/T_A \right)^{\frac{1}{2}}$ 
until $\alpha(\CT)$ reaches $\alpha_B$
around $\CT_B \sim  \Lambda_r^{3/2}/\Lambda_m^2 $.
In $\CT \gg \CT_B$, the matter dominates and the universe expands as 
$e^{\alpha(T)} \sim e^{\alpha_B}
\left( T/T_B \right)^{\frac{2}{3}}$.
We note that 
\eq{eq:smallE}
is indeed satisfied throughout the radiation-dominated era and afterward 
because
$\frac{C_ne^{n \alpha(\CT)}}{ E(\CT) e^{3 \alpha(\CT)}} 
\sim
\frac{C_ne^{n \alpha(\CT)}}{ C_n^{1/2} e^{3 \alpha(\CT) - \frac{6-n}{2}\alpha(\CT) }} 
\sim
\left( C_ne^{n \alpha(\CT)} \right)^{1/2}
\gg 1$ for $\CT \gg \CT_A$.
Therefore, the approximation used in 
\eq{psiaphigeneral}
is justified.
In these eras, the effective Hamiltonian is given by
\bqa
\hat H_{eff}(\CT)
=
\frac{2}{(6-n) \CT} 
\BL \hpa -
\sqrt{C_n} 
\hat \Pi
e^{\frac{n}{2} \ha}
\BR
\label{eq:Heff_general}
\eqa
to the leading order in $e^{-\alpha}$,
where 
$\hat \Pi$ 
is an operator that takes eigenvalue $s$ for 
$\Psi_{E,q}^{(s)}(\alpha,\phi)$.
In the regime where the WKB approximation is valid, $\hat \Pi \approx \hpa/|\hpa| $.
The effective Hamiltonian does not depend on $p_\phi$ to the leading order in $e^{-\alpha}$.

\subsection{
Dark-energy-dominated era}

Around time $\CT_C \sim \Lambda_m/\Lambda_1^{3/2}$,
the wavefunction becomes peaked at $\alpha_C$.
Beyond this size,
the dark energy dominates and 
\eq{eq:PsiEqn} becomes
\bqa
\Psi_{E,q}(\alpha,\phi) =
\frac{
e^{ \frac{3}{2} \alpha }
e^{i \left( q \phi 
\pm \eta(\alpha)
\mp E \xi(\alpha)
\right) }
}
{ \left[ \Lambda_0 e^{6\alpha} + \Lambda_1 e^{4\alpha} \right]^{1/4} },
\label{eq:PsiEqn3}
\eqa
where 
$\eta(\alpha)= 
\frac{e^{2\alpha}}{3}
\frac{
\Lambda_0^2 e^{4\alpha}
+ 3 \Lambda_0 \Lambda_1 e^{2\alpha}
+ 3 \Lambda_1^2 
}
{
\left( \Lambda_0 e^{2\alpha}+\Lambda_1 \right)^{3/2} + \Lambda_1^{3/2}
}$, 
$\xi(\alpha) =
\frac{1}{4 \sqrt{\Lambda_0}}
\log\left[
\frac{
\left( 
\sqrt{ \Lambda_0 e^{2\alpha}}
+
\sqrt{ \Lambda_0 e^{2\alpha} + \Lambda_1}
\right)^2
}
{\Lambda_1} \right]$.
The soft projection gives the time-dependent wavefunction,
\bqa
\Psi(T)
=
\frac{e^{ \frac{3}{2} \alpha }}{\sqrt{\CT}}
\sum_{s=\pm}
\int d\kphi ~
\Phi_s(0,\kphi )
\frac{
e^{
i \left( \kphi  \phi  
+  s \eta(\alpha)  
 \right) }
e^{ -\frac{\xi(\alpha)^2}{2 \CT^2 }}
}
{ \left[ \Lambda_0 e^{6\alpha} + \Lambda_1 e^{4\alpha} \right]^{1/4} }.
\label{psiaphiLambdac}
\eqa

In the first part of the dark-energy-dominated era,
$\Lambda_0$ is negligible,
and
\eq{psiaphiLambdac}
reduces 
to \eq{psiaphigeneral} 
with $n=4$ and $C_4=\Lambda_1$.
In this era, the universe expands as 
$e^{\alpha(\CT)} \sim  e^{\alpha_C} (T/T_C)$,
and
its unitary evolution is governed by \eq{eq:Heff_general} for $n=4$.

\begin{figure}
    \centering
    \subfigure[]{
\includegraphics[width=0.9\linewidth]{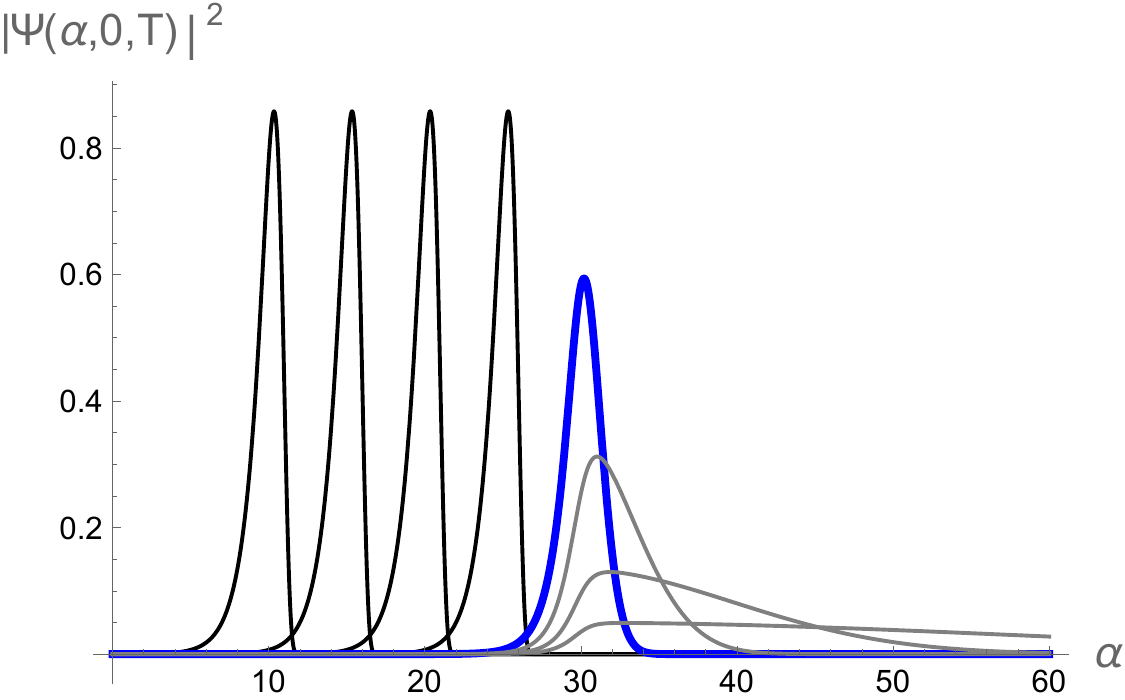}
\label{fig:3a}
}
\subfigure[]{
\includegraphics[width=0.9\linewidth]{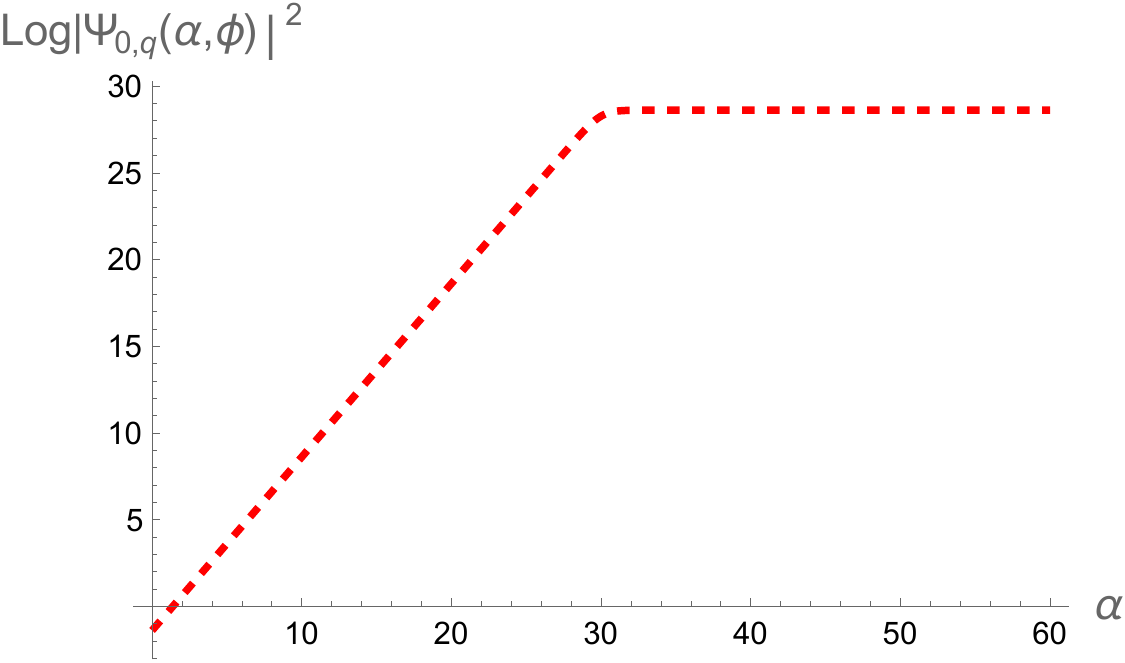} 
\label{fig:3b}
}
\caption{
(a)
The amplitude of the wavefunction
in the dark-energy-dominated era
with
$\Lambda_1=1$,
$\Lambda_0=e^{-60}$
and
$\Phi_+(0,q) = \delta(q)$,
$\Phi_-(0,q) = 0$.
Each curve represents
$|\Psi(\alpha,0,T)|^2$ as a function of $\alpha$
for $T=e^{10},e^{15},e^{20},e^{25},e^{30},e^{31},e^{32},e^{33}$ from left to right. 
The curve in the thick (blue) line is at $T^*=e^{30}$,
which marks the crossover from the $\Lambda_1$-dominated to the $\Lambda_0$-dominated evolution.
At the crossover time, the wavefunction is peaked around 
$\alpha^* = 30$.
For $T<T^*$, the peak of the wavefunction moves to larger values of $\alpha$ with increasing $\CT$.
Beyond $T^*$, the wavefunction gets broader in $\CT$ with the peak position fixed at $\alpha^*$.
(b) The logarithm of the amplitude of the gauge invariant wavefunction $\Psi_{E=0,q}(\alpha,\phi)$ in  
\eq{eq:PsiEqn3} for the same choice of parameters as in (a).
}
\label{fig:darkenergy}
\end{figure}

At $\CT^* = \Lambda_0^{-1/2}$,
the evolution crossovers 
to the $\Lambda_0$-dominated era and
the wavefunction is peaked around
 $\alpha^* \equiv \frac{1}{2} \log \frac{\Lambda_1}{\Lambda_0}$.
In $\CT \gg \CT^*$,
the form of the wavefunction becomes
qualitatively different.
In $\alpha  \ll \alpha^*$,
it is still described by 
\eq{psiaphigeneral} 
with $n=4$.
In $\alpha  \gg \alpha^*$, 
however,
the wavefunction becomes 
\bqa
&& \Psi(T)
=
\CT^{-1/2} \Lambda_{0}^{-\frac{1}{4}}  \times \nn &&
\sum_{s=\pm}
\int d\kphi ~
\Phi_s(0,\kphi )
e^{
i \left( \kphi  \phi  
+  s \frac{\sqrt{\Lambda_{0}}}{3} e^{3 \alpha } 
 \right) }
e^{ -\frac{\left(\alpha- 
\alpha^*
\right)^2}{8 \Lambda_{0} \CT^{2} }}.
\label{psiaphiLambdac4}
\eqa
As is shown in \fig{fig:3a},
the peak of the wavefunction is pinned at  $\alpha^*$,
and 
the width grows as $\Delta \alpha \sim \CT$ on the side of $\alpha>\alpha^*$.
While the expectation value of $e^\alpha$ grows exponentially in $T$,
the wavefunction acquires an increasingly large uncertainty of $\alpha$.
In this era,
the effective Hamiltonian, 
which can be written as
$\hat H_{eff}(\CT)
=
\frac{1}{\CT} \BL 
(\ha-\alpha^*) 
\left(\hpa
-
\sqrt{\Lambda_0 } 
e^{3\ha} 
\hat \Pi
\right)
\BR$ 
in $\alpha \gg \alpha^*$,
describes the broadening of the wavefunction.
Therefore, the semi-classical time evolution ends once the $\alpha$-independent cosmological constant dominates.
The change in the character of the time evolution in the $\Lambda_0$-dominated era can be understood from the profile of gauge-invariant wavefunction.
For 
$\alpha <  \alpha^*$, 
the gauge-invariant wavefunction 
$\Psi_{E=0,q}(\alpha,\phi)$ in 
\eq{eq:PsiEqn3} grows exponentially in $\alpha$ as is shown in \fig{fig:3b}.
If an initial wavefunction is localized in $\alpha < \alpha^*$, the projection $e^{-\frac{\CT^2}{2} \hat H^2}$ pushes the wavefunction to the region with larger amplitude to maximize the overlap, which gives rise to the directed semi-classical time evolution.
On the other hand, the amplitude of
$\Psi_{E=0,q}(\alpha,\phi)$ becomes flat in 
$\alpha >  \alpha^*$,
and the projection makes the wavefunction evolve diffusively.

We note that the underlying principle for our result is general beyond the minisuperspace model. 
Systems with configuration-dependent effective mass generally exhibit a directed unitary evolution under a stochastically induced wavefunction collapse 
(see Appendix \ref{app:particle}).

\section{Discussion}

In the present proposal,
the physical Hilbert space includes states that are not invariant under the temporal diffeomorphism and describe moments of time. 
Those instantaneous states evolve in time $T$ which is related to the violation of the Hamiltonian constraint through 
$T \sim 1/\sqrt{\langle \Psi(T) | \hat H^2 | \Psi(T) \rangle}$.
The time evolution can be viewed as one big measurement that causes a gauge non-invariant initial state to continuously collapse toward a gauge-invariant state.
Time flows toward the direction of reducing the violation of the Hamiltonian constraint.
When this scenario is applied to general relativity,
a directionality arises 
out of the wavefunction collapse 
due to the configuration-dependent effective mass.
The evolution becomes linear and unitary at large $T$ because of the universal manner in which the norm of the wavefunction changes under the collapse. 
At each instance, one can use the standard rule of quantum mechanics to compute probabilities.

The interpretation of wavefunction collapse as time evolution holds for general time  $T$.
We propose that the present universe corresponds to a state of a large but finite $T$.
This makes the present theory qualitatively different from other approaches that treat $\hat H$ as an exact gauge symmetry.
While the violation of the Hamiltonian constraint may be too small to be detected at present,
it may still be possible to measure it along with a signature of non-unitarity in the past 
from data with a large redshift 
because the cosmological data that reaches us today from distant objects is expected to exhibit larger fluctuations of $\hat H$ than that from nearer objects\footnote{I thank Latham Boyle for pointing this out.}.

Time $T$, which is related to the expectation value of $\hat H^2$, is a physical observable in our theory.
It is noted that $T$ is also proportional to the parameter time 
associated with a constant lapse for the initial value problem of general relativity:  
for matter with energy density $\rho \sim e^{-m \alpha}$
the scale factor evolves as $e^\alpha \sim T^{2/m}$, 
as is predicted for the parameter time with a constant lapse in general relativity.
Since $T$ is related to both $\langle \hat H^2 \rangle$ and the scale factor,
one can relate the latter two:
$\langle \hat H^2 \rangle \sim e^{-m \alpha}$.
Therefore,
$\Delta \alpha \equiv \int_{T_1}^{T_2} \frac{d \alpha}{dT} dT$, which can be inferred from the history of expansion between two moments in time, is related to the change in the violation of Hamiltonian constraint as
$\Delta \alpha \sim -\frac{1}{m} \Delta \log 
\langle \hat H^2 \rangle$.
%
If a measurement on our current universe gives a non-zero upper bound of $\hat H^2$, it does not rule out the possibility of $\hat H^2$ being zero.
On the other hand, if a measurement with an improved precision provides positive evidence for a non-zero $\hat H^2$, it falsifies the conventional description based on the exact temporal diffeomorphism. 
In the latter case, the relation between $\alpha$ and $\hat H^2$ can be used as an observable that verifies or falsifies our prediction.

The next step is to extend the present theory to the full general relativity.
If the Hamiltonian constraint is relaxed,
the full theory is expected to support one scalar mode, in addition to the two tensor graviton modes due to one fewer gauge constraint. 
For the tensor gravitons, it is expected that a unitarity emerges in the large $T$ limit.
This is because the tensor gravitons are excitations within the sub-Hilbert space
that satisfies the Hamiltonian constraint 
and that sub-Hilbert space is not affected by the projection. 
The overall scale factor is still expected to increase in $T$ due to the configuration-dependent effective mass, and 
the resulting unitary dynamics will describe the evolution of gravitons 
with respect to the increasing scale factor.
On the other hand, 
the extra scalar mode describes excitations outside the sub-Hilbert space with $\hat H=0$,
and is expected to be strongly damped.
It will be of great interest to confirm this explicitly 
and understand its physical consequences.

Relaxing the strict gauge symmetry to an approximate one is a drastic departure from the established theories for non-gravitational interactions in nature.
However, the temporal diffeomorphism is fundamentally different from the gauge symmetries of the standard model in that it is non-compact.
The non-compactness of the group makes it impossible to emerge as a gauge symmetry from a soft constraint.
If all gauge constraints in nature are enforced through soft constraints, the compact groups become gauge symmetries at low energies,
but the temporal gauge symmetry fails to emerge, giving us time.

The wavefunction collapse is not only an intrinsic part of time evolution\cite{
PhysRevD.34.470,
PhysRevA.42.78,
PhysRevA.40.1165,
Penrose:1996aa}
but is the very driving force.
However, the current scenario
does not necessarily exclude other forms of resolution for the measurement problem.
Through decoherence, 
for example,
one can, in principle, experience an additional effect of wavefunction collapse
within the unitary evolution that emerges in the late time limit.

\section*{Acknowledgments}

The research was supported by
the Natural Sciences and Engineering Research Council of
Canada.
Research at Perimeter Institute is supported in part by the Government of Canada through the Department of Innovation, Science and Economic Development Canada and by the Province of Ontario through the Ministry of Colleges and Universities.

\section*{DATA AVAILABILITY}

No data were created or analyzed in this study.

\bibliographystyle{unsrtnat}
\bibliography{references_high}


\appendix

\section{
Altered moments of time with a rotation of the basis for the clock variable
}
\label{app:txexample}

In this appendix, we illustrate how the notion of moments can be drastically altered upon a rotation of the basis used to measure the clock variable. 
As a simple example, let us consider a particle moving on the edge of the integer quantum Hall state\cite{PhysRevB.25.2185}.
The time-dependent Schrodinger equation reads 
$i \frac{\partial}{\partial t} \Psi = \hat p_x \Psi$,
where $\hat p_x 
 \equiv -i \frac{\partial}{\partial x}$ 
is the conjugate momentum of the position $x$ of the particle.
Because electrons move chirally on the quantum Hall edge, the Hamiltonian is simply proportional to the momentum.
The velocity has been set to be $1$.
This theory can be cast into a reparameterization-invariant form 
in which both $t$ and $x$ are treated as dynamical variables subject to a constraint,
$\hat H \Psi = 0$,
where
$\hat H = 
\hat p_t - \hat p_x$ is the self-adjoint operator that generates the reparameterization transformation.
Here,
$\hat p_t \equiv i \frac{\partial}{\partial t}$ is the conjugate momentum of $t$.
General gauge invariant wavefunctions
take the form of $\Psi(x,t) = f(x-t)$.
If we choose $t$ as our clock variable,
an instant is defined by measuring it.
Upon the projective measurement of the clock with outcome $t$, 
the probability for the outcome of the consequent $x$ measurement becomes 
$P(x|t)=|\Psi(x,t)|^2/\int dx' |\Psi(x',t)|^2
$.
This reproduces the predictions of the standard quantum mechanics.

 \begin{figure}[ht]
 \begin{center}
   \subfigure[]{
 \includegraphics[scale=0.3]{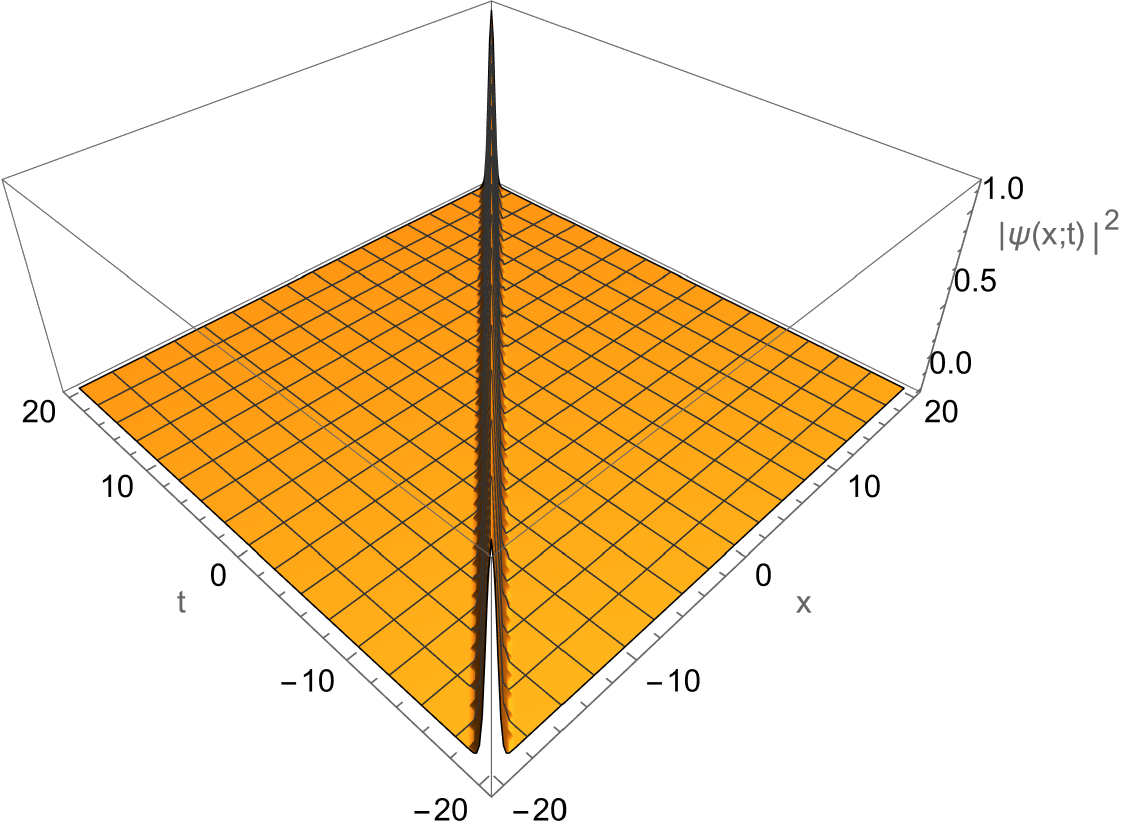} 
  \label{fig:t0}
 } 
  \hfill
    \subfigure[]{
 \includegraphics[scale=0.33]{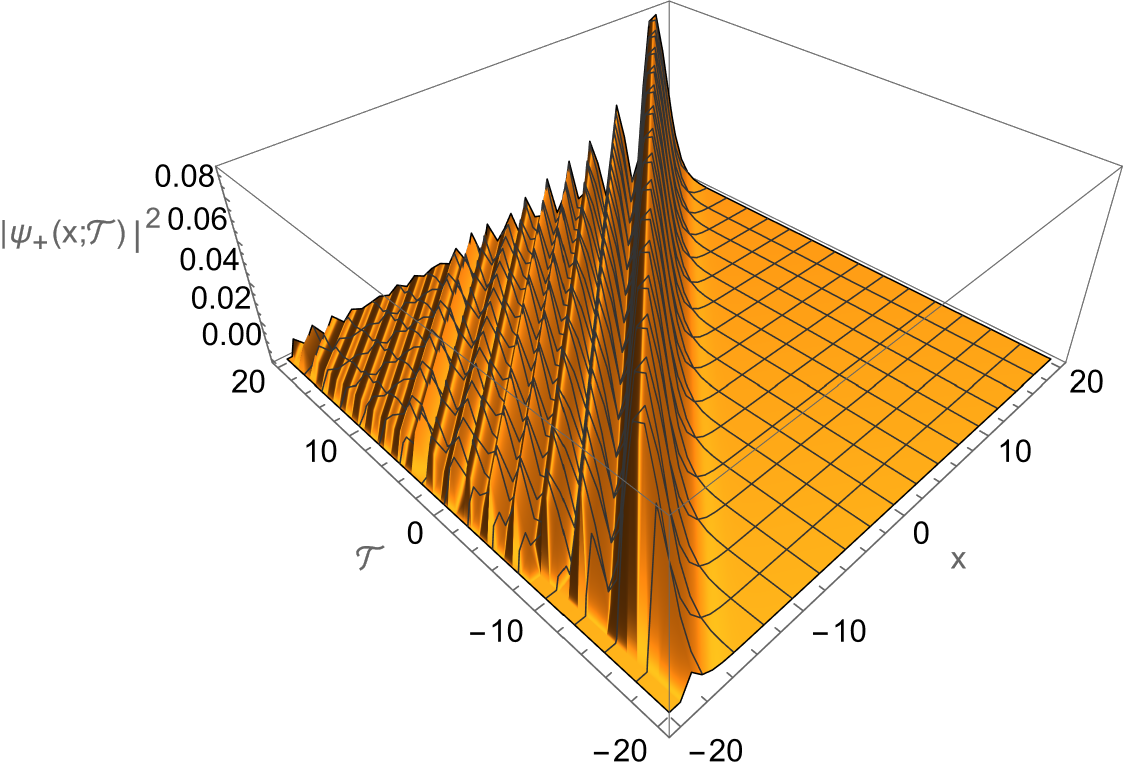} 
  \label{fig:T2}
 } 
 \end{center}
 \caption{
(a) The conditional probability for $x$ at time $t$ 
for the gauge invariant wavefunction 
$\Psi(x,t)
=$$ 
e^{
-\frac{\alpha^2}{2} 
(x-t)^2
+ i k_0 (x-t)
}
$ with $\alpha=2$ and $k_0=1$. 
It describes a wave-packet 
 well localized in $x$ propagating with speed $1$.
(b) 
The conditional probability for $x$ at an instant defined in basis 
$| \bt  \rangle_+$ with $\gamma=2$.
The state of $x$ at time $\bt$ is delocalized in $x<\bt$ because a moment of time in this basis includes the far past of the original basis.
 }
 \label{fig:psitT}
 \end{figure}

However, there is freedom to rotate the basis in which the projective measurement of the clock variable is performed to define a moment of time.
This is similar to the fact that a spin can be measured in any orientation.
After all, $t$ is a dynamical variable just like $x$ in this relational quantum mechanics.
Among many possible choices, we now use one alternative basis.
This choice has no particular significance other than it best illustrates how a moment of time defined in one basis mixes widely different moments of time defined in another basis.
Consider a new basis given by
\bqa
| \bt  \rangle_\pm = 
\int dt
~
\frac{1}{\gamma} Ai\left(\pm \frac{t-\bt}{\gamma}\right)
| t  \rangle,
\label{eq:basis2}
\eqa
where $Ai(t)$ is the Airy function and $\gamma$ is a positive constant.
$\frac{1}{\gamma} Ai\left(\frac{t-\bt}{\gamma}\right)$
$\left[ \frac{1}{\gamma} Ai\left(-\frac{t-\bt}{\gamma}\right) \right]$
is peaked around $t=\bt$  with 
its amplitude exponentially suppressed for  $t > \bt$  [$t < \bt$]
with width $\gamma$ 
but only power-law suppressed for 
$t < \bt$ [$t > \bt$].
The new basis satisfies
$_+\langle \bt' | \bt  \rangle_+
=
~ _-\langle \bt' | \bt  \rangle_-
= \delta(\bt-\bt')$.
Therefore, one may well define a moment of time from the projective measurement 
of the clock in the 
$| \bt  \rangle_+$
or
$| \bt  \rangle_-$ basis.
Upon the measurement 
of the clock in 
$| \bt  \rangle_\pm$,
the conditional probability of $x$ is 
controlled by an instantaneous wavefunction,
$
\Psi_\pm(x,\bt)  =
\int dt
~
\frac{1}{\gamma} Ai\left(\pm \frac{t-\bt}{\gamma}\right)
\Psi(x,t)$.
Since the Hamiltonian  is invariant under the unitary transformation that generates the basis change,
$\Psi_\pm(x,\bt)$ satisfies the same Schrodinger equation,
$i \frac{\partial}{\partial \bt}
\Psi_\pm(x,\bt)
=
\hat p_x
\Psi_\pm(x,\bt)
$.
At a moment of time defined in this new basis,
the system is in a linear superposition of states with vastly different $t$.
 This is illustrated in \fig{fig:psitT}.

\section{
Failed emergence of Coulomb phase from a soft non-compact gauge constraint}
\label{app:gauge_symmetry}

In models that exhibit emergent gauge symmetry,
the full Hilbert space of microscopic degrees of freedom includes states that do not satisfy Gauss's constraint.
Nonetheless, gauge theories can dynamically emerge at low energies in the presence of interactions that energetically penalize states that violate the constraint.
In this appendix, we review how this works for a compact group and discuss how it fails for the non-compact counterpart.

\subsection{U(1) group}

Here, we consider a lattice model where the pure U(1) gauge theory emerges at low energies.
Let $\hat \theta_{i,\mu}$ be the $U(1)$ rotor variable defined on link $(i,\mu)$ of the $d$-dimensional hyper-cubic lattice,
where $i$ is the site index
and $\mu=1,2,..,d$ denotes $d$ independent directions of links.
For links along $-\mu$ direction, we define
$\hat \theta_{i,-\mu} = - \hat \theta_{i-\mu,\mu}$.
$\hat n_{i,\mu}$ denotes the conjugate momentum of 
$\hat \theta_{i,\mu}$.
With $\theta_{i,\mu} \sim  \theta_{i,\mu} + 2\pi$,
$\hat n_{i,\mu}$ takes integer eigenvalues.
The Hamiltonian is written as
\bqa
\hat H = 
U \sum_i \hat Q_i^2
+
g \sum_{i,\mu} \hat n_{i,\mu}^2 
+ ...,
\label{eq:model3}
\eqa
where
$\hat Q_i  \equiv \sum_\mu (\hat n_{i \mu} - \hat n_{i-\mu,\mu})$
and $...$ denotes other terms.
The first two terms in the Hamiltonian respect the local $U(1)$ symmetry for every link,
but $...$ may partially or completely break the symmetry.
For example, we add
$\hat H_J = - 2 J \sum_{i,\mu} \cos(\hat \theta_{i,\mu})$ that breaks all internal symmetry.
We are interested in the low-energy spectrum of the theory
 in the limit that $U$ is larger than all other couplings.
If we view $\hat n_{i,\mu}$ as the electric flux in direction $\mu$, 
$\hat Q_i$ corresponds to the divergence of the electric field evaluated at site $i$.
The $U$-term in the Hamiltonian penalizes states that violate Gauss's constraint.
In the $U \rightarrow \infty$ limit, 
Gauss's constraint is strict, and 
states with finite energies only have closed loops of electric flux lines.

For a finite $U$, Gauss's constraint is not strictly enforced.
However, the gap between the low-energy sector with energy $E \ll U$ and the sector with $E \sim U$
guarantees that the low-energy Hilbert space  evolves adiabatically 
as $U$ is decreased from infinity to a finite value as long as $U$ is much larger than other couplings.
Therefore, there remains a one-to-one correspondence between states with $Q_i=0$
and the states with $E \ll U$ for a large enough $U$.
This guarantees that there exists a unitary transformation $\hat V$ that rotates the basis 
such that the Hamiltonian has no off-diagonal elements that mix the $Q_i=0$ sector and the rest.
In the rotated basis,
Gauss's law becomes an exact constraint within the low-energy Hilbert space with $E \ll U$.
Using the standard degenerate perturbation theory,
one can derive the pure U(1) gauge theory as the low-energy effective Hamiltonian,
\bqa
\hat V \hat H \hat V^\dagger  = 
g \sum_{i,\mu} \hat n_{i,\mu}^2 
- \sum_C t_C
\cos
\left( \sum_{ (i,\mu) \in C} \hat \theta_{i,\mu}
\right),
\label{eq:model32}
\eqa
where $t_C \sim J (J/U)^{L_C-1}$ 
with $L_C$ being the length of closed loop $C$.
For $g \ll t_\Box$, the gauge theory is in the deconfinement phase that supports $(d-1)$ gapless photons. 
The gaplessness of the photon is protected from small perturbations.
Therefore, 
the Coulomb phase emerges through the soft Gauss constraint for the U(1) group.

\subsection{$R$ group}

Now, we consider a non-compact counterpart of  \eq{eq:model3} by replacing
$\theta_{i,\mu}$ with 
a non-compact variable $\hat x_{i,\mu}$,
\bqa
\hat H = 
U \sum_i \hat Q_i^2
+g \sum_{i,\mu} \hat p_{i,\mu}^2 
+ ....
\label{eq:model4}
\eqa
Here, 
$\hat p_{i,\mu}$
denotes the conjugate momentum of $\hat x_{i,\mu}$.
Their eigenvalues can take any real number.
$\hat Q_i  \equiv \sum_\mu (\hat p_{i \mu} - \hat p_{i-\mu,\mu})$
is the generator of a local $R$ transformation at site $i$.
The symmetry-breaking perturbation, which is included in $...$, is written as 
$\hat H_J = J \sum_{i,\mu} \hat x_{i,\mu}^2$.
The question is whether the local $R$ symmetry emerges at a large but finite $U$ in the presence of such perturbations.
For simplicity, let us consider only $\hat H_J$ in the perturbation, 
which is enough for our purpose.
In this case, the theory 
 is quadratic and can be exactly solved.
In the Fourier space, we write
$\left(
\begin{array}{c}
 \hat x_{i,\mu} \\
\hat p_{i,\mu} 
\end{array}
\right)
= \frac{1}{\sqrt{L^d}} 
\sum_{m}
\sum_k
\left(
\begin{array}{c}
 \hat x_{k}^{(m)} \\
 \hat p_{k}^{(m)} 
\end{array}
 \right)
\varepsilon_{k,\mu}^{(m)}
~
e^{i r_i k}$,
where 
$k = \frac{2 \pi}{L}(l_1,..,l_d)$ 
with $l_i =  -L/2,.., L/2-1$ denotes discrete momenta that are compatible with the periodic boundary condition for the system with linear size $L$.
$\varepsilon_{k,\mu}^{(m)}$ with $m=1,..,d$ denotes the polarization of the $m$-th mode
with
$\varepsilon_{k,\mu}^{(m)*}
\varepsilon_{k,\mu}^{(n)}
=\delta_{m,n}$.
In terms of the Fourier mode,
the Hamiltonian becomes diagonal,
\bqa
\hat H =\sum_{k,m}
\left[
J \hat x_{k}^{(m)} \hat x_{-k}^{(m)}
+
V_{k,m} \hat p_{k}^{(m)} \hat p_{-k}^{(m)}
\right],
\eqa
where
$V_{k,1}= g + 4U 
\sum_\mu \sin^2 \frac{k_\mu}{2}
$
and
$V_{k,m \geq 2}= g$.
Here, $m=1$ represents 
the longitudinal mode with
$\varepsilon_{k,\mu}^{(1)} = \frac{
e^{\frac{i}{2}  k_\mu }  
\sin\left(\frac{k_\mu}{2}\right)
}{ \sqrt{ \sum_\nu \sin^2\left(\frac{k_\nu}{2}\right) } } 
$,
and $2 \leq m \leq d$ represent
$(d-1)$ transverse modes.
The energy dispersion of the mode is given by
$E_{k,m} = 2 \sqrt{J V_{k,m}}$.
It is noted that all excitations are gapped for any $J, g > 0$.
Therefore, there is no gapless photon.

The failed emergence of the Coulomb phase is a consequence of the fact that the states with $Q_i=0$ are in the middle of the spectrum with continuously varying $Q_i$.
Because there is no gap between the gauge invariant states and others, an arbitrarily small perturbation mixes states with different eigenvalues with an $O(1)$ weight.
It destroys the one-to-one correspondence between the gauge invariant states and the low-energy states 
for any non-zero $J/U$.
This can also be understood in terms of the ground state,
\bqa
\langle x | \psi_0 \rangle
=
e^{
-\frac{1}{2}
\sum_{k,m}
\sqrt{\frac{
J}{
V_{k,m}
}}
\left| x_{k}^{(m)} \right|^2
}.
\eqa
In the thermodynamic limit, the trace distance between the ground state
and the state obtained by applying a local $R$ transformation 
$e^{i y \hat Q_0}$ 
at the origin is
\bqa
d_{\psi_0}(y)
&=&
\sqrt{
1-
e^{
-2 y^2 
\int \frac{dk}{(2\pi)^d} 
\sqrt{\frac{
J}{
 g + 4U 
\sum_\mu \sin^2 \frac{k_\mu}{2}
}}
\sum_\nu \sin^2 \frac{k_\nu}{2}
}
}.
\label{eq:symmetry4}
\eqa
As expected, only the longitudinal modes contribute to the trace distance.
Due to the soft longitudinal mode,
there always exists $y$ for which \eq{eq:symmetry4} becomes $O(1)$
for any $J/U \neq 0$.

\section{Reduced FRW model}
\label{app:FRW}

In principle, we should treat all degrees of freedom on an equal footing.
Let us write the full Hamiltonian as
\bqa
\hat H = 
\BL
e^{-3 \ha} 
(-\hpa^2 +  \hpp^2 )
+ \hat h_X
\BR.
\label{H_FRW_full}
\eqa
Here, $X$ collectively represents all other degrees of freedom that include radiation, matter and other fields that source the dark energy
and $\hat h_X= h(\ha, \hat X, \hat p_{X})$ denotes the Hamiltonian that governs their dynamics.
Let $|X(\alpha) \rangle$ be an eigenstate of $\hat h_X$
with energy density $\rho_0(\alpha)$ at each $\alpha$:
$\hat h_X
|X(\alpha) \rangle
=
e^{3\alpha} \rho_0(\alpha)
|X(\alpha) \rangle
$.
Now, we consider a sub-Hilbert space defined by the projection operator,
\bqa
\hat {\cal P} =
\int d \alpha d\phi ~
| \alpha, \phi \rangle
\langle \alpha, \phi |,
\label{eq:ansatz}
\eqa
where 
$| \alpha, \phi \rangle 
\equiv 
|\alpha \rangle 
\otimes |\phi \rangle \otimes 
|X(\alpha) \rangle$.
For 
$|\Psi \rangle =  \int d \alpha d\phi ~ \Psi(\alpha,\phi)  |\alpha, \phi \rangle$,
the Hamiltonian projected to the sub-Hilbert space acts as 
\bqa
{\cal \hat P}\hat H |\Psi \rangle
= 
\int d \alpha d\phi ~
\left[ {\cal H} \Psi(\alpha,\phi) \right]
|\alpha, \phi \rangle,
\label{eq:ansatz2}
\eqa
where
\begin{widetext}
\bqa
\left[ {\cal H} \Psi(\alpha,\phi) \right]
&= &
\frac{1}{2} 
\Bigg\{
\Big[ e^{-3\alpha} \Big(
\partial_\alpha^2 
+ \langle X(\alpha) | \partial_\alpha^2 | X(\alpha) \rangle 
+2  \langle X(\alpha) | \partial_\alpha | X(\alpha) \rangle \partial_\alpha
- \partial_\phi^2  \Big) 
+ e^{3\alpha} \rho_0(\alpha)
\Big]
+ h.c. 
\Bigg\} 
\Psi(\alpha,\phi),
\label{eq:ansatz02}
\eqa
\end{widetext}
where $h.c.$ represents the Hermitian conjugate.
Without loss of generality, 
we can choose the phase of  $|X(\alpha) \rangle$ such that  $\langle X(\alpha)|  \partial_\alpha |X(\alpha) \rangle = 0 $
because $\alpha$ is non-compact.
With $\rho(\alpha) \equiv \rho_0(\alpha)
+
e^{-6\alpha}
\langle \Psi(\alpha) | \partial_\alpha^2 | \Psi(\alpha) \rangle
$,
we obtain the projected Hamiltonian in \eq{H_FRW},
\bqa
{\cal H}
=
\BL
e^{-3 \alpha} 
(\partial_\alpha^2 -  \partial_\phi^2 )
+ 
e^{3 \alpha} 
\rho(\alpha)
\BR,
\label{eq:calH}
\eqa
where $\rho(\alpha)$ behaves as an $\alpha$-dependent energy density contributed from $X$ degrees of freedom.

\section{Particle in a position-dependent effective mass}
\label{app:particle}

The main ingredient for the emergence of a directed unitary evolution under a stochastically induced wavefunction collapse is the configuration-dependent effective mass.
Therefore, our conclusion applies to a broader class of systems beyond the gravitational theory.
Here, we consider another example. 
A configuration-dependent effective mass can arise for quasiparticles in solids with nonuniform chemical compositions\cite{PhysRevB.27.7547} or strains\cite{AMORIM20161}. 
A particle in a nonuniform background is described by the Hamiltonian similar to \eq{H_FRW},
\bqa
\hat H
= 
\BL
\frac{1}{2m(\hat x)} \hat p^2
+ V(\hat x) \BR,
\eqa
where $m(x)$ and $V(x)$  represent the position-dependent mass
and potential, respectively.
In the WKB approximation, the eigenstate of energy $E$ can be written as
\bqa
\Psi_{E,s}(x) 
\sim 
\sqrt{
\frac{m(x)}{P_E(x)}
}
e^{i s \int dx P_E(x)},
\eqa
where
$P_{E}(x) 
=
\sqrt{
-2 m(x) V(x) + \left( \frac{m'(x)}{2m(x)} \right)^2 + 2 m(x) E  
}
$
and $s=\pm$ sets the sign of the momentum.
For simplicity, let us consider the case with
$m(x) = m_0 +  \gamma x^2$  and $V = V_0 <0$.
For initial wavefunction 
$\Psi_0(x) = \int dE \Phi_0(E) \Psi_{E,s}(x)$ with smooth $\Phi_0(E)$,
the gradual projection toward the sub-Hilbert space with zero energy 
through \eq{eq:mastereq2} 
results in 
\bqa
\Psi(x;T) 
\sim 
\Phi(0)
\sqrt{\frac{|x|}{T}} 
~
e^{
i s 
\sqrt{\frac{-\gamma V_0 x^2}{2}}  x
-\frac{1}{2T^2}
\left( \frac{\gamma}{-8V_0} \right)
x^{4}
}
\eqa
at large $T$.
In the large $T$ limit, the wavefunction evolves unitarily with increasing $T$, exhibiting peaks at $x \sim \sqrt{T}$
and
$-\sqrt{T}$.
The wavepacket centered at  $x \sim \sqrt{T}$ ($-\sqrt{T}$) drifts to the right (left) as $T$ increases.
Wavepackets always move toward the region with increasing effective mass irrespective of the sign of momentum $s$. 
This is because the preferred sign of the time step that maximizes the constructive interference under the stochastic sum in \eq{eq:averagedstate} is determined from the sign of momentum and the gradient of the mass. 
In $x>0$, $m'(x)>0$ and the wavepacket with a positive (negative) momentum preferably chooses positive (negative) time steps to move to the right irrespective of $s$. 
In $x<0$, $m'(x)<0$
the preferred sign of the time step becomes opposite to that of momentum so that the wavepacket always moves to the left.

\end{document}